# ICTs and Forced Migration: A Critical Discourse Review

Silvia Masiero
University of Oslo, Department of Informatics
silvima@ifi.uio.no

Mette von Deden
Loughborough University, School of Business and Economics
m.von-deden@lboro.ac.uk

**Abstract**

*The role of information and communication technologies (ICTs) in relation to issues of forced migration has come to the attention of the Information Systems (IS) literature. In this paper we review the interdisciplinary research on the topic, seeking to identify the main discourses in it. Three discourses – respectively centred on the migrant as a subject of inclusion, as an object of surveillance, and as an information seeker – emerge from the literature, offering important conceptual tools to map themes relating to ICTs and forced migration. Based on these discourses, we propose a different perspective – centred on the datafication of the migrant – which relates to existing data justice discourses and adds to the existing IS literature on the topic, offering a lens to further interpret the position of the forced migrant in relation to ICTs.*

## 1. Introduction

The Information Systems (IS) field has become increasingly interested in issues of forced migration. Such an interest has acquired various forms: themes of information and communication technologies (ICTs) and refugee integration have inspired conference panels in the Association of Information Systems, or AIS (AbuJarour et al., 2016, 2017), research agendas (AbuJarour et al., 2019) and journal publications (cf. Diaz Andrade & Doolin, 2016, 2018; AbuJarour et al., 2021; Madon & Schoemaker, 2021). Such an interest is inscribed in a shift of the IS field from "societal challenges" (Majchrzak et al., 2016) to issues of power and justice, which highlight the macro-political nature of challenges. Signalled by calls for papers centred on social justice (Aanestad et al., 2021) and the ethical and unintended consequences of digitalisation and platformisation (Rossi et al., 2020), such a shift towards power and justice is gradually but substantially reshaping the IS field.

The IS literature is, however, not alone in tackling themes of ICTs and forced migration. Bearing in mind the importance for IS scholarship to learn from other fields (Walsham, 2012), it needs to be noted that the overlap of ICTs and forced migration is observed in many disciplines, making it challenging to map the domains across which the theme is tackled. From human geography to development, media and communications and critical data studies, aspects of ICTs and forced migration have been captured across perspectives, geographies and theories. Such an interdisciplinary attention makes it important to avail a guiding light on existing knowledge on the topic, to gain a clear picture of themes that have been undertaken and perspectives that have been devised.

This paper is an effort to produce such a guiding light. Starting from an IS literature angle, the paper reviews published research on ICTs and forced migration, focusing especially on what IS scholarship can learn from research conducted in other fields. Using discourse mapping as a scoping technique, we identify the main discourses in this emerging literature, relying on the notion of *discourse* as a set of central assumptions and ideas on a topic (Avgerou, 2010). In doing so, we integrate IS perspectives on ICTs and forced migration with other literatures, mapping the landscape on a topic that is raising interest across disciplinary boundaries.

We find three discourses in the literature on ICTs and forced migration. These are centred, respectively, on the migrant as a subject of inclusion, as an object of surveillance and as an information seeker, and find significant overlaps in the conceptualisation of the migrant in relation to new technologies. The perspectives are mutually integrated: the themes found in each of them resonate with one another. But the three discourses also remark important differences across fields: in particular, themes where the IS literature finds limited presence are those of information seeking and, especially, surveillance and injustice, finding instead a prevalence of research focusing on inclusion and integration.

On this basis, the paper brings a novel perspective to the IS literature. We rely on the three discourses in point to offer a notion of the migrant as a datafied subject, for whom datafication conditions essential aspects of existence. The perspective relies on Mann's (2018) theory of datafication and integrates it in the notion of data justice (Taylor, 2017), showing how datafication of the migrant is a fundamental completion to existing IS literature. Doing so, we integrate IS perspectives with one related to the role data play in migrant lives, further exploring the migrant's position in relation to new technologies.

This paper is structured as follows. We first present the results of a structured literature search of IS journals, eliciting core themes in IS research on ICTs and forced migration. We then conduct a discourse mapping of the interdisciplinary literature on the topic: this identifies three central discourses, offering three routes to explore the theme. We then propose the theoretical foundations of a view of the migrant as a datafied subject, showing the importance of notions of data justice to integrate the IS literature. Doing so, we offer a critical discourse review that is a means to take stock of the literature, but also to introduce a datafying vision in it.

## 2. Themes in IS literature on ICTs and forced migration

Our considerations start from a mapping of research on ICTs and forced migration in the IS field. Similar efforts (Alencar, 2020) have been conducted in interdisciplinary terms, but a landscape review of IS research on the topic is so far missing. We have started by a mapping for relevant keywords (migrant; migrants; migration; refugee; refugees; displaced; displacement) of the AIS Senior Scholars Basket of Journals, on the grounds that these journals are internationally recognised as representative of the IS field. To these we have added the journals Communications of the AIS, due to its representativeness of panels and discussions occurring in forums of the Association for Information Systems (AIS), and the Electronic Journal of Information Systems in Developing Countries, due to its engagement with topics relevant to socio-economic development such as forced migration. While we had included the journal Information Technology for Development in a first review, our reading of search results from the journal did not reveal an openly IS approach (with the exception of Schreieck et al., 2017), and we have thus moved all other relevant papers from this journal to the interdisciplinary literature section of the paper.

Our search found a total of 44 papers, 13 of which engage dynamics of forced migration in relation to ICTs. While different across theories and empirical contexts, such papers reveal three common threads that characterise the emerging IS literature on ICTs and forced migration:

1. *A socio-technical approach to the matter*. Of the 13 relevant papers in our corpus, 11 are empirical studies and two are reports from panel sessions at AIS conferences, one of which also presents some interview data. The papers span across contexts (countries in which studies are set include Malawi, Uganda, Germany, New Zealand, South Africa, Singapore, Canada and the United States); they do, however, find common ground in the sociotechnical approach to the topic, which views technology as closely intertwined with the human experience of forced migration. Whether the focus is on digital identity platforms for refugees (Madon & Schoemaker, 2021), on ICT inclusion initiatives (Diaz Andrade & Doolin, 2016, 2018) or on information practices among migrants (Gomez, 2016; Gomez & Vannini, 2017), technology is seen as embedded in human experience, and examined in its ability to condition, through integration or hurdle generation, the migrant's journey. Characteristic of the IS field, the socio-technical approach (Hirschheim et al., 1991) is deeply reflected in IS research on forced migration, and gives it a precise identity within the broader literature on the topic.

2. *Empirical work centred on resettled refugees*. Forced migration happens through different stages of transition and resettlement, which differ significantly from each other (Newell et al., 2016). Both transition and resettlement can be difficult and traumatic, with transition being life-threatening and ridden with danger of deportation and police violence (Latonero & Kift, 2018). Against this backdrop, all but one of the empirical papers in our corpus focus on migrants that have either resettled, or are in the process of doing so. Only one paper, by Madon and Schoemaker (2021), focuses instead on the transition context of a refugee camp: another paper researching a refugee camp setting (Brown et al., 2021) depicts refugees in the Dzaleka camp of Malawi as already resettled, conceiving the camp as a stable place of living. While presenting a rich picture of resettlement, in the IS literature we find a paucity of studies of the transition context, one in which information practices proliferate (Newell et al., 2016) and that could therefore provide an important space for the generation of IS knowledge on the lived experience of migrants.

3. *Focus on social inclusion*. Across the papers in point, a common matrix emerges in the framing of the role of ICTs in migrants' lives. This is a focus on ICTs as a route to social inclusion, variously but positively participating in resettlement. Examples range from a governmental initiative for ICT-based inclusion of refugees in New Zealand (Diaz Andrade & Doolin, 2016, 2018), to an information platform for refugees in Germany (Schreieck et al., 2017) and the use of ICTs for entrepreneurship by refugees in Malawi (Brown et al., 2021). This does not mean that IS research is oblivious of ICT-induced injustices: for example, Madon and Schoemaker (2021) detail the eruption of domestic violence in refugee households, as a consequence of men's reaction to the establishment of women as household heads in the digital identity system. At the same time, a positive framing of ICTs as embedded in social inclusion pervades IS papers, further contributing to its distinctive identity in the interdisciplinary literature.

A socio-technical approach, empirical work centred on resettled refugees, and a focus on social inclusion characterise the IS body of literature on forced migration. On the one hand, a thin body of literature of just 13 papers offers only preliminary considerations on a body of knowledge that, as sessions in AIS conferences and new calls for papers in IS journals reveal, is expanding rapidly. On the other, the pervasiveness of the three threads already affords establishing a specific IS identity to forced migration research, an identity that emerges when placed in a broader literature context.

## 3. Discourses on ICTs and forced migration

Over time, scholars from multiple disciplines have increasingly dealt with ICTs and forced migration. The emergence, noted by Alencar (2020), of a field called *digital migration studies* is revealing of such an increase in scholarly attention, with multiple foci in terms of technologies and migrant populations.

The notion of *discourse* – conceived, with Avgerou (2010), as a set of assumptions and ideas on a topic – offers a guiding light in mapping a complex literature, whose richness makes it challenging to identify a circumscribed list of disciplines of reference.

We have hence resorted to mapping the discourses encountered in our extensive read of sources on ICTs and forced migration. Started by extending our keyword search across disciplines, our literature search brought us to fields as diverse as refugee studies, media and communications, human geography, development studies, and diverse areas of sociology and anthropology. Discourse mapping has led us to distinguish three bodies of assumptions and ideas depicting the migrant, respectively, as a subject of inclusion, as an object of surveillance, and as an information seeker. All three discourses are illustrated here.

### 3.1. The migrant as a subject of inclusion

As noted above, the social inclusion of migrants through ICTs is a common focus in the IS literature. With a study of New Zealand, Diaz Andrade and Doolin (2016) show how resettled refugees experience ICTs as a route to entrance in a new society, illustrating five capabilities – participation in an information society, effective communication, understanding of a new society, social connection and expression of a cultural identity – that refugees achieve through ICTs. In a further study of the same context (Diaz Andrade & Doolin, 2018), three types of ICT-mediated information practices – sustaining support networks, maintaining transnational ties, and expressing cultural identity – are found to be constitutive of refugee inclusion. Foci of the IS literature, such as digital platform properties, are also related to refugee inclusion: for example, Schreieck et al. (2017) show how the properties of a refugee information platform are constitutive of inclusion, and Madon and Schoemaker (2021) show how digital identity platforms can be designed to improve refugee management.

However, a discourse centering the ICT-connected migrant (Diminescu, 2008) as subject of inclusion is found well beyond the boundaries of the IS literature. First, sociological understandings are diffused on the role of technology in the formation of belongingness to a new society: for example, Schaub's (2012) study of Congolese migrants in Morocco underlines the cruciality of mobile communication in providing inclusive infrastructures for the making of a new life. Aricat's (2015) study of migrant laborers in Singapore, and Vancea and Olivera's (2013) work on migrant women in Catalonia, all detail the sense of security attached by the migrant to communication devices, fostering a much needed belongingness to a new society. Papers from a social inclusion discourse maintain a positive undertone, where ICTs are seen in their light as participants to the resettlement of the migrant, supporting establishment in the country of settlement.

### 3.2. The migrant as an object of surveillance

Based in diverse literatures, a surveillance discourse illuminates other sides of technology. These reside in the surveillance practices resulting from migrants' use of ICTs: while potentially helpful in navigating the needs of migration, digital technologies afford profiling of the migrant, which is dangerous both in transition and in resettlement. In his analysis of three EU migration databases, Broeders (2007) studies how, with the development of a network of databases, member states are seeking to establish control on settled irregular migrants. In this way, technologies function to exclude illegal migrants from the key institutions of society, supporting police authorities in facilitating deportation.

Building on the same argument, Pelizza (2020) explores the relation between surveillance and the interoperability established across migration databases. With a shift occurred in July 2015, the Eurodac database of asylum seekers was made interoperable with national police authorities in the EU, making it possible for national police forces to resort to EU databases of asylum seekers when investigating crime (Pelizza, 2020: 263). Pelizza's (2020) notion of *alterity processing* illuminates the link between database interoperability and surveillance, whose consequences are made more acute in contexts of risk for the migrant (Milan et al., 2020; Trauttmansdorff, 2021). Information practices emerge among irregular migrants in response to surveillance: in a study of undocumented migration across the US-Mexico border, Newell et al. (2016) illuminate the use of border disturbance technologies to prevent migrant deaths in the desert, but also to map and avoid the risk of violence and deportation.

Mann (2018) reflects on the origin of the problem: the amount of data available on previously invisible populations (for example, data from phone activity captured by telecommunication towers) increases their readability, scaling use of such data in the name of "development" (Taylor & Broeders, 2015). Justifying such practices are, Mann (2018) continues, the humanitarian grounds of data extraction, which provide moral and ethical grounds to surveillant mappings of migrant populations. Backed by humanitarian principles, social injustices such as the conditionality of social protection to biometric profiling of refugees (Iazzolino, 2021) find detailed mention in the surveillance discourse.

As this stream of literature reveals, surveillance may result in forms of punishment that affect not only the migrant, but also their extended networks made accessible to police forces through ICTs. Monroy (2021) reports on how, as a punishment for irregular entry, border guards in Europe have resorted to breaking migrants' mobile phones, while police forces capture these to gain information on networks and routes for irregular migrant entrance. The surveillance discourse shows, along the same line, how risk of violence and deportation affects migrants' behaviour, resulting in shying away from requesting essential services such as hospital care for the fear of being deported (Milan et al., 2020). Research from this discourse shows the importance of a perspective that unpacks the surveillance potential of technologies, as well as the countersurveillance routes against them.

### 3.3. The migrant as information seeker

In a third discourse the migrant, who experiences complex information needs (Borkert et al., 2018), is seen in the light of their use of ICTs to manage such needs. This is true for migrants in transition, such as the crossers of the US-Mexico border studied by Newell et al., (2016), but also for resettled migrants, such as the Syrian and Iraqi migrants into Italy studied by Harney (2013). Migrants in this study use the mobile phone to solve the uncertainties of daily life, such as visa issues and the search of a job, similarly mitigating risk through connectivity (Harney, 2013: 243).

Connected to information seeking behaviour is the behaviour of the migrant that, deprived from the vicinity of family and social networks, uses ICTs to communicate with them. The discourse shows especially how intimate connections with families and friends in home countries are grounded on the basis of technological means. This is so for female migrant workers in Singapore (Thomas & Lim, 2010), displaced people in Turkey refugee camps (Smets, 2019) and Filipino workers in the UK, for whom the mobile is the means of communication with children left behind (Madianou & Miller, 2011). When communication is oriented to finding work opportunities, it opens up studies of migrant platform workers, with the risks and enhanced vulnerabilities that platform work entails (Howson et al., 2020).

Studies of the digital diaspora, which explore how ICTs have impacted the lived experience of migration (cf. Leurs & Ponzanesi, 2018; Candidatu et al., 2019; Smets, 2019; Leurs, 2019); also fall into an information seeking discourse. These studies offer an anthropological examination of how ICTs affect the experience of migration and displacement: the establishment of a connected presence (Diminescu,

2008) poses the migrant at the intersection between societies, inviting questions on how digitality affects the migrant's identity (Leurs, 2019). On the one hand the societal visibility coming with connectivity reinforces pre-existing identity; on the other, the mobile phone acts as a bridge that inscribes that identity in the new societies, generating new forms of identity conscience (Leurs & Ponzanesi, 2018).

In sum, for this stream of literature the migrant is, in the first place, an information seeker that uses technologies for their own needs of information and communication. This problematises the vision of migrants as passive (Borkert et al., 2018) and reconceives them as active subjects of processes of information seeking and communication. Risks are numerous for both transition and resettled migrants, and the wrong information may lead to police violence, deportation and even death. In managing risks, the migrant arises as an agent that leverages technologies to carve their life routes.

| Discourse | Assumptions | Themes | Vision of technology |
|---|---|---|---|
| Social Inclusion | ICTs are a means to social inclusion of migrants, supporting their integration in the host country<br><br>Properties of ICTs (e.g. design properties of digital identity platforms) make the social inclusion of migrants possible | Government initiatives for ICT-based migrant inclusion; information platforms; refugee entrepreneurship; digital identity platforms; humanitarian platforms for social assistance and protection | Positive: ICT as a force for good |
| Surveillance | ICTs are a means to undue surveillance of migrants, which puts them in danger during both transition and resettlement<br><br>Interoperability of databases is leveraged to profile migrants across borders and enable deportation | Migrant databases; interoperability; law and policy on ICTs and migration; police violence; enforcement of borders; data injustice on irregular and undocumented migrants and refugees | Negative: ICT as enabling oppression and violence |
| Information Seeking | ICTs are a means for the migrant to satisfy their urgent and complex information needs | Mobile connectivity in relation to information seeking; communication patterns with home and host countries; | Ambivalent: ICT as capable to enable information seeking, but also to expose to danger of deportation |

| | Properties of ICTs afford both information search and communication with relevant counterparts (family left behind; friends; informant entities) | digital diaspora; cultural identity established through digital technologies | |

*Table 1.     Discourses on ICTs and Forced Migration*

Table 1 shows the core features of each of the three discourses. While IS perspectives are mainly centred on social inclusion, some have elements of information seeking (Gomez, 2016; Gomez & Vannini, 2017) and recognise the limitations of the inclusion discourse. As the table shows, discourses centred on surveillance and information seeking present important complementarities with social inclusion, complementarities that the IS literature is well-positioned to examine.

### 4. Broadening the IS literature: A datafying perspective

The exercise conducted in this paper has led to the mapping of a diverse literature on ICTs and forced migration. Examining the three discourses found in the literature, we have noted the resonance of themes across them and the different assumptions that each discourse displays. Transcending the IS field, we have embraced a variety of perspectives that further remarks the broadening of themes and views that the IS literature can undertake.

Multiple themes have emerged across discourses, and the interdisciplinarity of the literature is the reason for the plurality of visions proposed on them. A thematic thread, however, emerges above others: data – be they demographic, biometric, related to entitlement to income or refugee status – are central in the life of the migrant. From identification at the border to the reception of social protection, data are a constant marker of the migrants' experience, which all three discourses bring to light under different sets of assumptions.

Datafication – seen, with Cukier & Mayer-Schoenberger (2014), as the conversion of existing processes into data – is therefore a strong force in migrant lives. As Mann (2018) illustrates, it is the human being (and, at large, the migrant population) to be converted into data, so to become amenable through visualisation and administration by machine readability. Datafication affords core aspects of the migrants' lived experience: service provision, information seeking, but also police surveillance and cross-border profiling. Relatedly, the three discourses offer different visions on the consequences of the datafication of migrants:

- In the social inclusion discourse, datafication makes it possible to assist previously unseen and, thus, unassisted populations. For example, Madon and Schoemaker (2021) show how the conversion of human individuals (refugees in the Bidi Bidi camp in Uganda) into digital data affords the ability of the United Nations High Commissioner for Refugees (UNHCR) to match them with their entitlements and hence provide them with food and assistance. While recognising some injustices, the overarching view is one of datafication as a route to enabling effective social assistance.

- In the surveillance discourse, datafication is instrumental to the profiling that poses serious dangers for migrants during transition and resettlement. For example, Milan et al. (2020) study the condition of undocumented migrants during global health emergencies, showing the coupling of identifiability – necessary to access life-saving medical care – and policing, which may result in deportation. As a consequence of interoperability between databases, the datafied migrant who requires medical care can be flagged as irregular through the database, and is hence induced to refrain for requesting an assistance that could turn into life-threatening. Centred on profiling and policing, the surveillance discourse engages datafication as a route to unjust practices over migrants, to be dissected and known to generate countersurveillant mechanisms in response.

- Finally, in the information seeking discourse datafication is double-edged: on the one hand, having one's own data stored in a database simplifies the research process. For example, migrants into Italy studied by Harney (2013) witnessed simplification of visa procedures due to existing data on them. On the other hand, exposing information may substantially complicate the seeking process: the migrants into the US studied by Newell et al. (2016) declare avoiding the use of vulnerable mobile devices, due to the risk that information in them may lead to capture and deportation. In this discourse, datafication is hence capable of supporting virtuous practices – such as information seeking – and dangerous ones, such as revelation of information whose sharing puts the migrant in peril.

All three discourses show that seeing the migrant as datafied sheds important light on a complex field. While different across fields, the theory of datafication is seen here as a route to draw attention to core aspects of migrants' existence, on which the IS literature can be significantly broadened.

Two implications emerge from this theoretical view. First, the notion of justice is also adapted to a datafied context: Taylor (2017) shows that just as the physical world needs principles of social justice, a datafied world needs a *data justice* that means fairness in visualisation, representation and treatment of people through their digital data. Migrant populations, previously largely invisible through data, are now increasingly datafied: this implies the need of a new type of justice for them, to protect them in the ways they are visualised, represented and treated through data. If a social justice is to be conceived, this should be also reflected in the treatment of data and in the use made of them by public authorities.

Secondly, the notion of data justice also enables exploring the range of injustices that the datafied world affords on the individual. Many of these affect migrants' situation: undue exposure of data may lead to attack, deportation or deadly violence (Coppi et al., 2021). From an informational perspective, information on migrant data treatment is often opaque, and biometric profiling is exchanged for aid that should be provided as a universal human right (Iazzolino, 2021). The surveillance discourse fits in here once again: justified by the need for policing, the interoperability of databases resulted in large-scale victimhood, caused by the link between asylum seeking and police intervention (Pelizza, 2021). Within this scenario, the notion of data injustice offers an important conceptual tool for the IS literature.

## 5. Conclusion

This paper has sought to understand what the nascent, but promising, IS literature on ICTs and forced migration can learn from other fields. Our discourse mapping found discourses centred on social inclusion, surveillance and information seeking. While focused on social inclusion, the IS literature has a lot to learn: notions of datafication and data justice are pivotal to build a nuanced understanding of the topic. In proposing these notions, we hope to have offered a route to expand and structure the upcoming IS literature on ICTs and forced migration.

**References**

Avison, D. E. and Fitzgerald, G. (1995). *Information systems development: Methodologies, techniques and tools*, 2nd Edition. London: McGraw-Hill.

Aanestad, M., Kankanhalli, A., Maruping, L., Pang, M., & Ram, S. (2021). Special Issue on Digital Technologies and Social Justice. MISQ Call for Papers, https://misq.org/skin/frontend/default/misq/pdf/CurrentCalls/SI_DigitalTechnologies.pdf.

AbuJarour, S., Ajjan, H., Fedorowicz, J., & Köster, A. (2021). ICT Support for Refugees and Undocumented Immigrants. *Communications of the AIS*, 48(1), 40-52.

AbuJarour, S., Wiesche, M., Andrade, A. D., Fedorowicz, J., Krasnova, H., Olbrich, S. & Venkatesh, V. (2019). ICT-enabled refugee integration: A research agenda. *Communications of the AIS*, 44(1), 874-891.

AbuJarour, S. A., Krasnova, H., Diaz Andrade, A., Olbrich, S., Tan, C. W., Urquhart, C., & Wiesche, M. (2017, June). Empowering refugees with technology: Best practices and research agenda. *European Conference on Information Systems*, Guimarães, 5-10 June 2017.

AbuJarour, S., Krasnova, H., Wenninger, H., Fedorowicz, J., Olbrich, S., Tan, C. W., & Urquhart, C. (2016). Leveraging Technology for Refugee Integration: How Can We Help? International Conference of Information Systems (ICIS), Dublin, 15-18 December 2016.

Alencar, A. (2020). Mobile communication and refugees: An analytical review of academic literature. *Sociology Compass*, 14(8), 1-13.

Aricat, R. G. (2015). Mobile Ecosystems Among Low-Skilled Migrants in Singapore: An Investigation into Mobile Usage Practices. *The Electronic Journal of Information Systems in Developing Countries*, 68(1), 1-15.

Avgerou, C. (2010). Discourses on ICT and development. *Information Technologies and International Development*, 6(3), 1-18.

Borkert, M., Fisher, K. E., & Yafi, E. (2018). The best, the worst, and the hardest to find: How people, mobiles, and social media connect migrants in (to) Europe. *Social Media+ Society*, *4*(1), 1-13.

Broeders, D. (2007). The new digital borders of Europe: EU databases and the surveillance of irregular migrants. *International Sociology*, 22(1), 71-92.

Brown, S., Saxena, D., & Wall, P. J. The role of information and communications technology in refugee entrepreneurship: A critical realist case study. *The Electronic Journal of Information Systems in Developing Countries,* 1-20.

Candidatu, L., Leurs, K., & Ponzanesi, S. (2019). Digital diasporas: Beyond the buzzword: toward a relational understanding of mobility and connectivity. *The Handbook of Diasporas, Media, and Culture*, 31-47.

Coppi, G., Jimenez, R. M., & Kyriazi, S. (2021). Explicability of humanitarian AI: A matter of principles. *Journal of International Humanitarian Action*, published online 7 October 2021.

Cukier, K., & Mayer-Schoenberger, V. (2014). *The rise of big data: How it's changing the way we think about the world*. New York: Princeton University Press.

Diaz Andrade, A., & Doolin, B. (2019). Temporal enactment of resettled refugees' ICT mediated information practices. *Information Systems Journal*, 29(1), 145-174.

Diaz Andrade, A., & Doolin, B. (2016). Information and communication technology and the social inclusion of refugees. *MIS Quarterly*, 40(2), 405-416.

Diminescu, D. (2008). The connected migrant: an epistemological manifesto. *Social science information*, 47(4), 565-579.

Gomez, R., & Vannini, S. (2017). Notions of home and sense of belonging in the context of migration in a journey through participatory photography. *The Electronic Journal of Information Systems in Developing Countries*, 78(1), 1-46.

Gomez, R. (2016). Vulnerability and information practices among (undocumented) Latino migrants. *The Electronic Journal of Information Systems in Developing Countries*, 75(1), 1-43.

Harney, N. (2013). Precarity, affect and problem solving with mobile phones by asylum seekers, refugees and migrants in Naples, Italy. *Journal of Refugee Studies*, 26(4), 541-557.

Hirschheim, R., Klein, H. K., & Newman, M. (1991). Information systems development as social action: Theoretical perspective and practice. *Omega*, 19(6), 587-608.

Howson, K., Ustek-Spilda, F., Grohmann, R., Salem, N., Carelli, R., Abs, D., Salvagni, J., Graham, M., Balbornoz, M.B., Chavez, H., Arriagada, A., & Bonhomme, M. (2020). 'Just because you don't see your boss, doesn't mean you don't have a boss': COVID-19 and Gig Worker Strikes across Latin America. *International Union Rights*, 27(3), 20-28.

Iazzolino, G. (2021). Infrastructure of compassionate repression: Making sense of biometrics in Kakuma refugee camp. *Information Technology for Development*, 27(1), 111-128.

Latonero, M., & Kift, P. (2018). On digital passages and borders: Refugees and the new infrastructure for movement and control. *Social Media+ Society*, 11(1).

Leurs, K. (2019). Transnational connectivity and the affective paradoxes of digital care labour: Exploring how young refugees technologically mediate co-presence. *European Journal of Communication*, 34(6), 641-649.

Leurs, K., & Ponzanesi, S. (2018). Connected migrants: Encapsulation and cosmopolitanization. *Popular Communication*, 16(1), 4-20.

Madianou, M., & Miller, D. (2011). Mobile phone parenting: Reconfiguring relationships between Filipina migrant mothers and their left-behind children. *New media & Society*, 13(3), 457-470.

Madon, S., & Schoemaker, E. (2021). Digital identity as a platform for improving refugee management. *Information Systems Journal*, 31(6), 929-953.

Majchrzak, A., Markus, M. L., & Wareham, J. (2016). Designing for digital transformation: Lessons for information systems research from the study of ICT and societal challenges. *MIS Quarterly*, 40(2), 267-277.

Mann, L. (2018). Left to other peoples' devices? A political economy perspective on the big data revolution in development. *Development and Change*, 49(1), 3-36.

Milan, S., Pelizza, A. and Lausberg, Y. (2020). Making migrants visible to COVID-19 counting: the dilemma, *Open Democracy*. Available at https://www.opendemocracy.net/en/can-europe-make-it/making-migrants-visible-covid-19-counting-dilemma/

Monroy, M. (2021). Frontex and Europol: how refugees are tracked digitally. https://digit.site36.net/2021/10/25/frontex-and-europol-how-refugees-are-tracked-digitally/

Newell, B. C., Gomez, R., & Guajardo, V. E. (2016). Information seeking, technology use, and vulnerability among migrants at the United States–Mexico border. *The Information Society*, *32*(3), 176-191.

Pelizza, A. (2021), Identification as translation: The art of choosing the right spokespersons at the securitized border, *Social Studies of Science:* 1-25.

Pelizza, A. (2020). Processing alterity, enacting Europe: Migrant registration and identification as co-construction of individuals and polities. *Science, Technology, & Human Values*, *45*(2), 262-288.

Rossi, M., Cheung, C., Sarker, S., & Thatcher, J. (2020). Special Issue on Ethical and Unintended Consequences of Digitalisation and Platformisation. Journal of Information Technology Call for Papers, https://journals.sagepub.com/pb-assets/cmscontent/JIN/JIT%20CFP%20SI%20Ethical%20Issues%20Digitalization.pdf.

Schaub, M. L. (2012). Lines across the desert: mobile phone use and mobility in the context of trans-Saharan migration. *Information Technology for Development*, *18*(2), 126-144.

Schreieck, M., Wiesche, M., & Krcmar, H. (2017). Governing nonprofit platform ecosystems–an information platform for refugees. *Information Technology for Development*, 23(3), 618-643.

Smets, K. (2019). Media and immobility: The affective and symbolic immobility of forced migrants. *European Journal of Communication*, 34(6), 650-660.

Taylor, L. (2017). What is data justice? The case for connecting digital rights and freedoms globally. *Big Data & Society*, 4(2), 1-14.

Taylor, L., & Broeders, D. (2015). In the name of Development: Power, profit and the datafication of the global South. *Geoforum*, 64, 229-237.

Thomas, M., & Lim, S. S. (2010). Migrant workers' use of ICTs for interpersonal communication–The experience of female domestic workers in Singapore. In *E-seminar Working Paper: Migrant workers' use of ICTs for interpersonal communication*.

Trauttmansdorff, P. (2021). Making biometric borders interoperable. A necessary Political Fiction? Presented at Biometrics on the Move 4S Annual Meeting, Society for the Social Study of Science, 8 October 2021.

Vancea, M., & Olivera, N. (2013). E-migrant women in Catalonia: Mobile phone use and maintenance of family relationships. *Gender, Technology and Development*, 17(2), 179-203.

Walsham, G. (2012). Are we making a better world with ICTs? Reflections on a future agenda for the IS field. *Journal of Information Technology*, 27(2), 87-93.